\documentclass[twocolumn]{aastex61}

\usepackage{xcolor}
\usepackage{amsmath}
\usepackage{rotating}

\definecolor{maroon}{rgb}{0.5, 0.0, 0.0}
\definecolor{dark-green}{rgb}{0.05, 0.5, 0.06}
\definecolor{dark-blue}{rgb}{0.0, 0.0, 0.5}

\received{February 20, 2018}
\revised{May 22, 2018}
\accepted{June 29, 2018}
\published{August 8, 2018}
\submitjournal{ApJ}

\shorttitle{3-phase evolution: the magnetic field}
\shortauthors{Heinemann et al.}

\begin{document}

\title{3-Phase Evolution of a Coronal Hole, Part II: The magnetic field}

\correspondingauthor{Stephan G. Heinemann}
\email{stephan.heinemann@hmail.at}

\author{Stephan G. Heinemann}
\affil{University of Graz, Institute of Physics, Universit\"atsplatz 5, 8010 Graz, Austria }

\author{Stefan J. Hofmeister}
\affil{University of Graz, Institute of Physics, Universit\"atsplatz 5, 8010 Graz, Austria }

\author{Astrid M. Veronig}
\affil{University of Graz, Institute of Physics, Universit\"atsplatz 5, 8010 Graz, Austria }

\author{Manuela Temmer}
\affil{University of Graz, Institute of Physics, Universit\"atsplatz 5, 8010 Graz, Austria }

\begin{abstract}
We investigate the magnetic characteristics of a persistent coronal hole (CH) extracted from EUV imagery using HMI filtergrams over the timerange February 2012 $-$ October 2012. The magnetic field, its distribution as well as the magnetic fine structure in form of flux tubes (FT) are analyzed in different evolutionary states of the CH. We find a strong linear correlation between the magnetic properties (e.g. signed/unsigned magnetic field strength) and area of the CH. As such, the evolutionary pattern in the magnetic field clearly follows the three-phase evolution (growing, maximum and decaying phase) as found from EUV data (Part I). This evolutionary process is most likely driven by strong FTs with a mean magnetic field strength exceeding $50$~G. During the maximum phase they entail up to $72\% $ of the total signed magnetic flux of the CH, but only cover up to $3.9\%$ of the total CH area, whereas during the growing and decaying phase, strong FTs entail $54-60\%$ of the signed magnetic flux and cover around $1-2\%$ of the CHs total area.
We conclude that small scale-structures of strong unipolar magnetic field are the fundamental building blocks of a CH and govern its evolution.
\end{abstract}


\section{Introduction} \label{sec:intro}
Coronal Holes (CHs) are large-scale structures in the solar corona consisting primarily of open magnetic field presumably rooted in the photosphere and are the major source of high speed solar wind. The open magnetic field is usually dominated by one magnetic polarity which is caused by a continuing imbalance in the local magnetic flux emergence \citep{1982levine,1996wang}. 
Looking into the fine-structure of the CH magnetic field, it is found that in local minima of flux emergence, i.e.\,between photospheric granulation cells, magnetic elements are accumulating to small scale unipolar structures called flux tubes (FT), network bright points or solar filigree \citep{1973dunn,2001berger,cranmer2009}. They can be observed in line-of-sight photospheric magnetograms as regions with areas of $ <$ few $10^{6} $km$^{2}$ which entail the majority of the signed magnetic flux coming from a CH, and therefore may be considered as CH "footpoints" \citep[e.g.,][]{2005tu,2017hofmeister}. 

In the higher atmospheric layers those small scale magnetic structures coincide mostly with the edges and nodes of the chromospheric network cells and expand rapidly in the higher chromosphere to merge to a near homogeneous vertical magnetic field in the corona \citep{1976gabriel,1986dowdy}. Funnels are open magnetic structures connecting the chromosphere with the solar corona \citep[][ and references therein]{2000hackenberg}. These so-called magnetic funnels have been suggested as the source regions of the high speed solar wind streams (HSS) within CHs \citep{1999hassler,2005wiegelmann,2005tu}. 

However, not all of the magnetic flux in CHs is unipolar, but a significant amount of closed field lines exists within its boundaries. \citet{2004wiegelmann} investigated the difference between CHs and the quiet sun. It was found that, despite a relative high amount of signed magnetic flux ($77\pm14\%$), closed magnetic loops still exist. The average height of those loops is found to be lower in CHs than in the quiet sun. This may be related to the funnel-like expansion of the FTs into the corona which form canopy-like regions between the funnels. The magnetic field as well as magnetic funnels of CHs have been studied and modeled especially in relation to the solar wind acceleration mechanism in low solar heights \citep[e.g.,][]{1982levine,wang90,gosling96,2009wang, 2017hofmeister} but details on the evolution of the magnetic field as well as the evolution of FTs in the context of CH evolution are still missing. As these intrinsic CH magnetic field properties shape not only the local field within the CH but also the global solar magnetic field \citep{2002bilenko,2013petrie,2014wiegelmann_review,2016bilenko}, a detailed analysis on the evolution of CHs together with its underlying magnetic field is of high interest. 
%

For investigating the photospheric magnetic field encompassed by the CH region, the boundaries of CHs need to be extracted. This is usually performed from hot coronal emission lines in the EUV and X-ray wavelength range. Owing to their low plasma density and temperature compared to the surrounding corona, at these wavelengths CHs are observed as dark structures and can be extracted with algorithms that are based on intensity thresholds \citep[eg. ][]{schwenn06,2009krista,2012rotter,2014reiss,2017hofmeister}. 

In the first part of this study \citep[hereafter referred to as paper~I][]{2018heinemann_paperI} the evolution of a long-lived (10 solar rotations) and low latitude CH was investigated using combined EUV image data and in-situ measurements from three different viewpoints covering $360\arcdeg$ of the heliosphere. With the usage of the two STEREO satellites \citep[Solar TErrestrial RElations Observatories;][]{2008kaiser_STEREO} the CH could be seamlessly tracked over its entire lifespan from which we derived a 3-phase evolution (growing, maximum, decaying) of the CH. These phases were most prominently revealed in the CH area evolution but were also obtained from multiple other parameters (e.g.\,intensity, associated solar wind speed). In addition to the area, intensity and parameters of the associated high speed stream, also the rotational and latitudinal motion of the CH has been analyzed in paper I.

The evolutionary pattern found in paper~I is most likely related to the underlying magnetic field and the presented study will give a better understanding of the magnetic evolution. Limited to Earth-view, we investigate the same long-lived low-latitude CH that was studied in paper~I, using data from the Heliospheric and Magnetic Imager \citep[HMI,][]{2012schou_HMI,2016couvidat_HMI} on-board of the Solar Dynamics Observatory \citep[SDO,][]{2012pesnell_SDO}. We analyze the change in the global CH magnetic field and its distribution as well as its fine structure in form of FTs and relate it to the evolutionary structure of the CH as derived in paper~I.

\section{Data and Methods} \label{sec:methods}

\subsection{Data} \label{subsec:data}
The HMI/SDO instrument acquires velocity, magnetic field and spectral line measurements of the solar photosphere from narrow-band filtergrams of six wavelengths centered on the spectral line of Fe \textsc{i} (6173~\AA). Sequences are obtained at a rate of 45s and 135s with an image size of 4096$\times$4096 pixel and a angular resolution of $0.5\arcsec$ \citep{2016couvidat_HMI}. For this study Line of sight (LoS) magnetograms were used for the magnetic field measurements. The 720s LoS data was used due to the lower photon noise of $\sim3$ G measured near the center of the solar disk and because of the higher signal-to-noise ratio at lower field strengths in comparison to vector magnetograms. 

To minimize projection effects, LoS magnetic field data were analyzed for each rotation of the CH around the time of the central meridian passage (CMP) of the CHs center of mass (CoM). This covers a spatial range of $\pm10\arcdeg$ longitude from the central meridian and is related to a time window of roughly $\pm18$ hours around the time of the central meridian passage of the CoM (assuming that the magnetic field does not change substantially over that time range). For each rotation during that time window, data are taken at a 1 hour cadence downloaded from the Joint Science Operations Center (JSOC). This results in a dataset of $\sim 25-30$ images per CMP. We calculate each parameter from each image and then average over each set to reduce noise and short term variations. In total, this gives 10 data points for studying the magnetic field evolution of the CH.


\subsection{Data Reduction and CH Extraction} \label{subsec:pdata-prep}
Basic data reduction was applied using the SolarSoft Suite of the Interactive Data Language (SSW-IDL). The images where prepped to level 1.5 and bad images (e.g. saturation or high noise) were removed. 

Using LoS magnetic field data we assumed a radial magnetic field which was corrected by applying a pixel-wise correction:
\begin{equation}
B_{\mathrm{i,corr}}=\dfrac{B_{\mathrm{i}}}{\cos (\alpha_{\mathrm{i}})},
\label{eq:radial-mag}
\end{equation}
with $B_{\mathrm{corr,i}}$ being the corrected value of each pixel of the magnetic field map, $B_{\mathrm{i}}$ the uncorrected one and $\alpha_{\mathrm{i}}$ the respective angular distance from the solar disk center.

To analyze the magnetic field underlying the CH, we apply CH masks extracted from EUV images to the co-registered photospheric magnetic field maps. The CH boundaries were extracted from EUV 193\AA\ images from the Atmospheric Imaging Assembly \citep[AIA,][]{2012lemen_AIA} by applying an intensity based threshold method ($35\%$ of the median intensity of the solar disk; for more details see paper~I). 

The projection-corrected area was calculated from the binary CH maps under assumption of a spherical sun: 
\begin{equation}
A_{\mathrm{i,corr}}=\dfrac{A_{\mathrm{i}}}{\cos(\alpha_{\mathrm{i}})},\label{eq:spherical-correction}
\end{equation}
with $A_{\mathrm{i}}$ being the area per pixel and $\alpha_{\mathrm{i}}$ the angle towards he center of the solar disk. From this we were able to calculate the size of the CH by adding up the corrected areas for all CH pixel:
\begin{equation}\label{eq:area}
A=\sum_{\mathrm{i}}^{\mathrm{N}}A_{\mathrm{i,corr}}.
\end{equation}
The area is given in square kilometers, $\mathrm{km^{2}}$. 

Similar as for the magnetic field, we average the calculated area in 1 hour cadence over a time window of $\pm18$ hours around the time of the CMP of the CoM (see Section 2.5 in paper~I). The error bars shown in Figures ~\ref{fig:3-mag-area-plot}, \ref{fig:flux}, \ref{fig:ft-number}, \ref{fig:ft-number-area} and \ref{fig:FT-flux-prop} are the $1 \sigma$ standard deviations from the mean values.

\subsection{Analysis of the Global Magnetic Field within the Coronal Hole} \label{subsec:global-prop}

The parameters of the global CH magnetic field, as extracted from the EUV mask, include the magnetic field strength, the magnetic flux and the magnetic field distribution with its moments (mean, variance, skewness, and kurtosis).

The mean (signed) magnetic field strength $\bar{B}$, in Gauss [G] was calculated as 
\begin{equation}\label{eq:meanmag}
\bar{B}=\dfrac{1}{N}\sum_{\mathrm{i}}^{\mathrm{N}}B_{\mathrm{i,corr}},
\end{equation}
where i represents the i-th pixel of the CH and $N$ is the total number of all pixel within the CH. Correspondingly, the unsigned magnetic field strength is calculated as
\begin{equation}\label{eq:abs-meanmag}
\bar{B}_{\mathrm{us}}=\dfrac{1}{N}\sum_{\mathrm{i}}^{\mathrm{N}}|B_{\mathrm{i,corr}}|.
\end{equation}

From the magnetic field strength and the area of the structure we can derive the magnetic flux, $\Phi$. The magnetic flux is given in Maxwell [Mx] and can be divided into signed (or \textit{open}) and unsigned (or \textit{total}) flux. The signed magnetic flux $\Phi_{\mathrm{s}}$ is calculated as the net flux through the respective area: 
\begin{equation}\label{eq:sflux}
\Phi_{\mathrm{s}}=\sum_{\mathrm{i}}^{\mathrm{N}}(B_{\mathrm{i,corr}}~A_{\mathrm{i,corr}}),
\end{equation}
with $B_{\mathrm{i}}$ and $A_{\mathrm{i}}$ the magnetic field strength and area for each pixel respectively. The unsigned magnetic flux $\Phi_{\mathrm{us}}$ is calculated as the total flux through the respective area: 
\begin{equation}\label{eq:usflux}
\Phi_{\mathrm{us}}=\sum_{\mathrm{i}}^{\mathrm{N}}(|B_{\mathrm{i,corr}}|~A_{\mathrm{i,corr}}).
\end{equation}

From the signed and unsigned magnetic flux we can define the magnetic flux balance $R_{\Phi}$ of a CH as
\begin{equation}\label{eq:fluxbalance}
R_{\Phi}=\dfrac{|\Phi_{\mathrm{s}}|}{\Phi_{\mathrm{us}}}.
\end{equation}
The signed flux divided by the unsigned flux of a CH can be seen as a measure of the flux that is not balanced within the CH and this provides a measure of the open flux percentage. 


\subsection{Flux Tube Extraction and Analysis} \label{subsec:ex}

To analyze the fine structure within the CH that is composed of FTs, i.e.\, small scale unipolar structures, we use a simple threshold based method that is applied on the magnetograms with a threshold value of $\pm20$~G. All structures with pixels above this threshold and containing at least 3 pixels where extracted as FTs. The value of $\pm20$~G was chosen because structures extracted with this and with higher thresholds show unipolar properties, whereas structures that have been extracted with a lower field threshold, may also contain pixels with opposite polarity \citep[see also ][]{2017hofmeister}. The value for the average unipolarities  for all FTs extracted is $0.99~[+0.01~-0.02]$, which gives support for the choice of the threshold of 20~G. The unipolaritiy is calculated using Eq.~\ref{eq:fluxbalance} as described in Section~\ref{subsec:ex}, which, for values around 1, indicates that all pixels of the structure (e.g.\,FT) are of the same polarity.

The total number of extracted FTs is called FT ensemble. This ensemble is then sorted into three categories based on the absolute value of the mean magnetic field strength of each FT structure$|\bar{\mathrm{B}}_{FT}|$, which is calculated as
\begin{equation}
|\bar{B}_{\mathrm{FT}}|=\dfrac{1}{N}\sum_{\mathrm{i}}^{\mathrm{N}}|B_{\mathrm{i,corr}}|,
\end{equation}
where i represents the i-th pixel of the FT and $N$ is the total number of pixels that make up the FT. 

\begin{sidewaysfigure*}
\centering \includegraphics[width=1\textheight , angle=0]{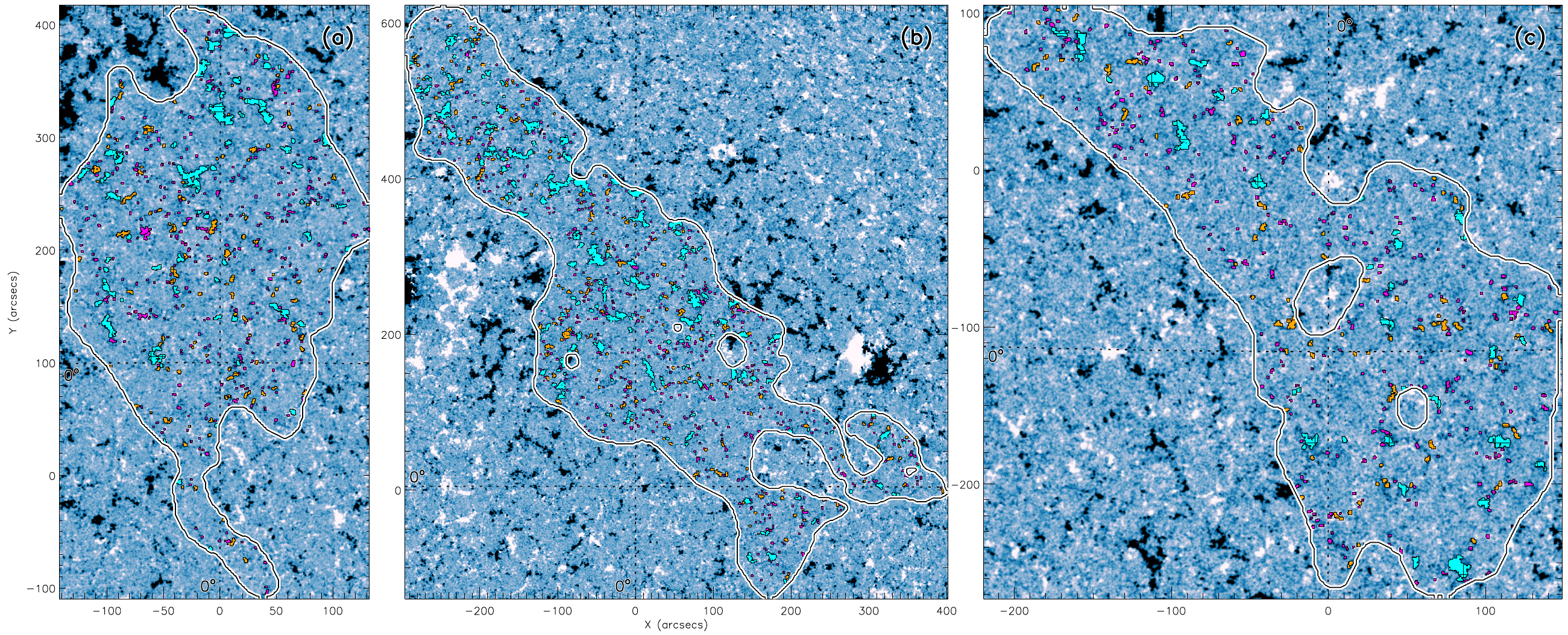}
\caption{Photospheric magnetograms, scaled to $\pm 20$G with the CH boundary overlayed (black) and with contours of FTs of different categories. The cyan contours show the strong FTs, the orange the medium FTs and the magenta the weak FTs. The panels represent different stages of the CH evolution (a: growing phase, April 09, 2012; b: maximum phase, June 03, 2012; c: decaying phase, August 22, 2012). Note that each image has a different axis-scaling, representing different zoom levels of the magnetiogram.}\label{fig:FT-example}
\end{sidewaysfigure*}

The first category has the lowest magnetic field strengths and are called \textit{weak} FTs. If the absolute value $|\bar{\mathrm{B}}_{FT}|$ of a FT is between $20$G and $35$G it is placed into the pool of weak flux tubes. The next category are the \textit{medium} FTs within an interval of the absolute value of the mean magnetic field strength from $35$G to $50$G. All FTs that exceed a mean magnetic field strength of $\pm50$G are called \textit{strong} FTs. Figure~\ref{fig:FT-example} shows three snapshots of the CH magnetic field (one of each phase) with the FTs highlighted. The \textit{strong} FTs are shown in cyan, the \textit{medium} FTs in orange and the \textit{weak} FTs in magenta. The FTs seem to be aligned along the magnetic network \citep{1976gabriel,1986dowdy} and a change in the relative abundance of the FTs during the CH evolution can be seen. 


The area of each FT is calculated using:
\begin{equation} \label{FTarea}
A_{\mathrm{FT}}=\sum_{\mathrm{i}}^{\mathrm{N}}A_{\mathrm{i,corr}},
\end{equation}
with $A_{\mathrm{corr,i}}$ being the area of each FT pixel corrected for a spherical sun (Eq.~\ref{eq:spherical-correction}) and N the number of pixels of the FT. The flux for each FT was calculated using Equations~\ref{eq:sflux} and~\ref{eq:usflux}. By summing over all FTs of one category in one image, we can calculate the FT property  of the total CH, e.g. $ \Phi_{\mathrm{FT,s}} $ is the signed flux coming from all (weak, medium or strong) FTs within the CH.

Lastly we can define the FT proportions, the area proportion and the flux proportion. These ratios show how much of a certain category of FTs contributes to the respective parameter of the total CH in terms of flux or area:
\begin{equation}\label{FTratio}
\begin{gathered}
r_{\Phi} = |\dfrac{\Phi_{\mathrm{FT}}}{\Phi_{\mathrm{CH}}}|\\
r_{\mathrm{A}} = \dfrac{A_{\mathrm{FT}}}{A_{\mathrm{CH}}}.
\end{gathered}
\end{equation}

\startlongtable
\begin{deluxetable}{ll|l}
\tablecaption{Overview of Parameters Defined in Section~\ref{sec:methods} \label{tab:methods}}
\tablehead{
\colhead{Parameter} & \colhead{Definition\tablenotemark{a,b}} & \colhead{Description}
}
\startdata
$A$ & $=\sum_{\mathrm{i}} A_{\mathrm{i}}$ & Area. \\
$\bar{B}$& $=\tfrac{1}{N}\sum_{\mathrm{i}} B_{\mathrm{i}}$& Signed mean magnetic  \\
 & & field strength. \\
$\bar{B}_{\mathrm{us}}$& $=\tfrac{1}{N}\sum_{\mathrm{i}} |B_{\mathrm{i}}|$& Unsigned mean magnetic \\
 & & field strength. \\
$\Phi_{\mathrm{s}}$ & $=\sum_{\mathrm{i}} (B_{\mathrm{i}}~A_{\mathrm{i}})$ & Signed magnetic flux. \\
$\Phi_{\mathrm{us}}$ & $=\sum_{\mathrm{i}} (|B_{\mathrm{i}}|~A_{\mathrm{i}})$ & Unsigned magnetic flux. \\
$R_{\Phi}$ & $=|\tfrac{\Phi_{\mathrm{s}}}{\Phi_{\mathrm{us}}}|$ &Flux Balance, ratio of signed \\
 & & to unsigned magnetic flux. \\
$r_{\Phi} $&$= \tfrac{|\Phi_{\mathrm{FT}}|}{\Phi_{\mathrm{CH}}}$ &Flux Ratio, proportion of flux \\
 & & from FTs to the CH flux. \\
$r_{\mathrm{A}} $&$= |\tfrac{A_{\mathrm{FT}}}{A_{\mathrm{CH}}}|$ &Area Ratio, proportion of area \\
 & & from FTs to the CH area. \\ \hline \hline
\enddata
\tablenotetext{a}{Note that $A_{\mathrm{i}}$ and $B_{\mathrm{i}}$ represent the corresponding \\ corrected versions $A_{\mathrm{i,corr}}$ and $B_{\mathrm{i,corr}}$}
\tablenotetext{b}{Coronal hole and flux tube properties are denoted with the \\ subscripts CH and FT respectively.}
\end{deluxetable}

The correlation between the various extracted parameters is calculated using the Pearson correlation coefficient and the Spearman correlation coefficient. To consider the significance of the relation for a low number of data points, we apply a bootstrapping algorithm \citep{efron93_bootstrap,efron1979_bootstrap} on the dataset with over $10^{6}$ repetitions (replacement for each repetition is taken from the initial set, with each data point in this subset coming from a Gaussian distribution of itself plus its standard deviation). The correlation coefficients are calculated from the derived subsets and given as mean values of all repetitions. The confidence intervals (CI) for the correlation coefficients ($90\%,~95\%,~99\%$) were calculated using the respective quantiles. A summary of all results and statistical parameters is given in Table~\ref{tab:corr} in the Appendix. 


\section{Results on Magnetic Coronal Hole Properties} \label{sec:global_res}
\begin{figure*}
 \centering %
  \includegraphics[width=0.82\linewidth]{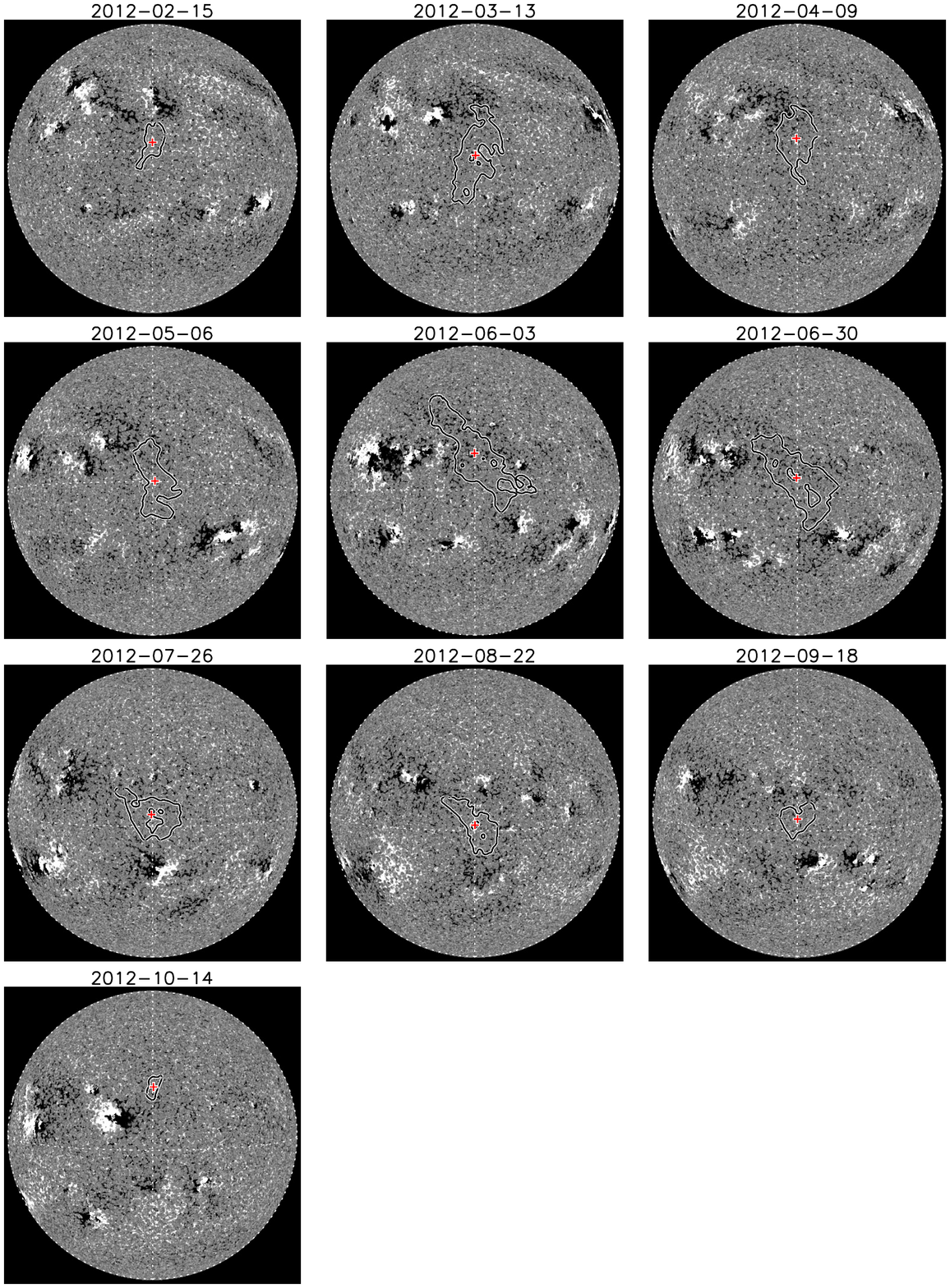}
\caption{Evolution of the LoS magnetic field below the CH during its lifetime over 10 solar rotations. HMI full disk LoS magnetograms are shown with black contours representing the CH boundaries extracted from the AIA images and the red cross showing the CoM of the CH. }\label{fig:evo-plot}
 \end{figure*}

In this section, we present the results of the evolution of the magnetic field properties of the CH under study. Snapshots of the evolution over the entire lifespan of the CH is shown in Figure~\ref{fig:evo-plot} with the full disk magnetograms and CH contours overlayed in black.

\subsection{Area and Magnetic Field Strength} \label{subsec:res:3-phase}
\begin{figure*}
\centering \includegraphics[width=1\linewidth,angle=0]{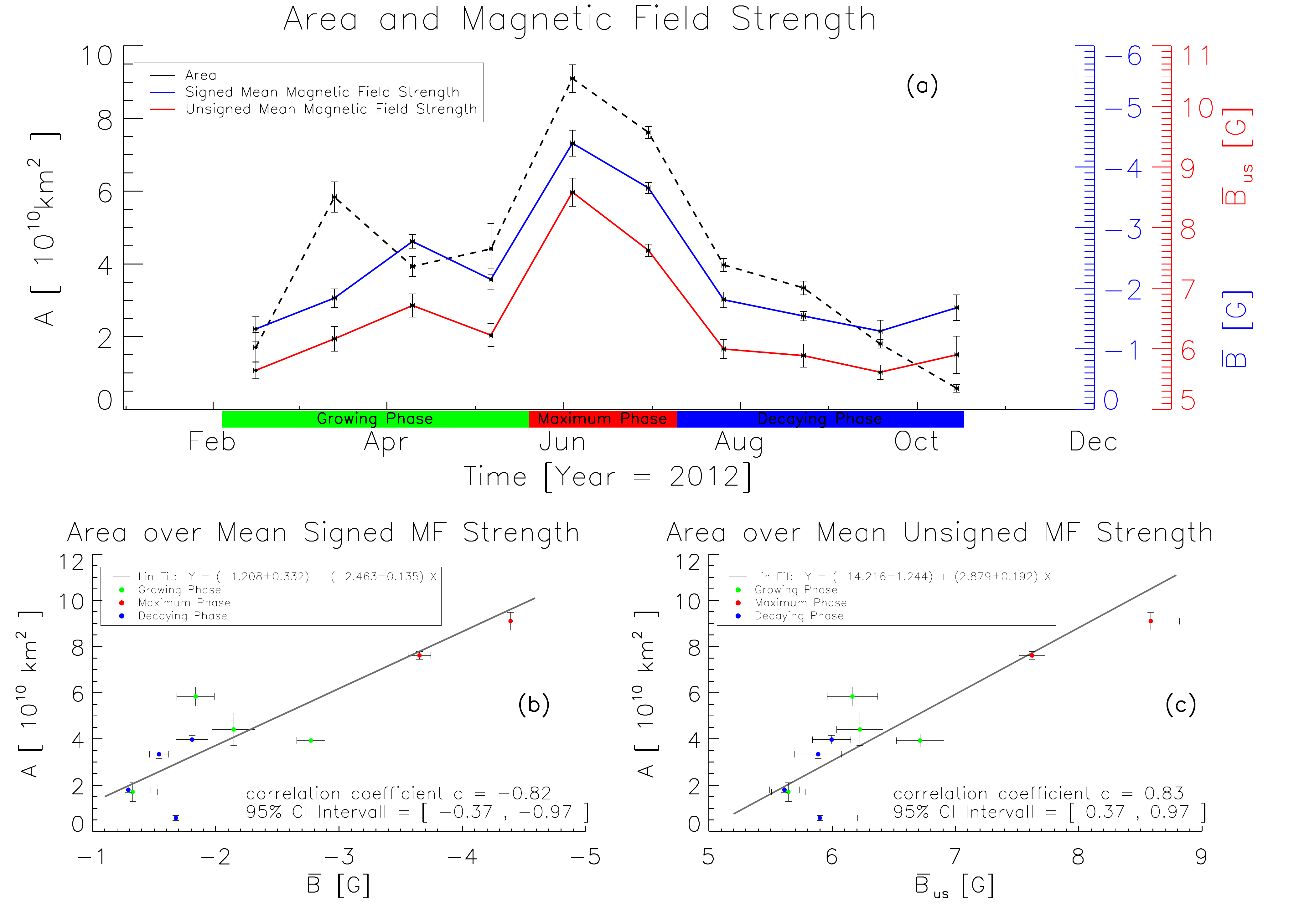}
\caption{(a): Evolution of the area (dashed, black), the signed mean magnetic field strength (red) and the unsigned mean magnetic field strength (blue) of the CH. (b): Correlation between signed mean magnetic field strength and area. (c): Correlation between unsigned mean magnetic field strength and area. The colored data point (green, red and blue) show their affiliation with an evolutionary phase (growing, maximum and decaying).}\label{fig:3-mag-area-plot}
 \end{figure*}

In the following, we take up the result of paper~I where we show the 3-phase evolution of the CH as derived from the CH area evolution (and other parameters like the intensity and the in-situ solar wind peak velocity of the associated HSS). Figure~\ref{fig:3-mag-area-plot}(a) shows the CH area evolution as observed from Earth-view together with the evolution of the mean magnetic field strength of the CH over its lifetime. From the CH area (dashed black line) we clearly obtain the 3-phase evolutionary pattern. In the growing phase, lasting from February 4, 2012 until May 13, 2012, we see a first peak in the deprojected area at $ \sim 6\cdot10^{10}$ km$^{2}$ followed by a fast decline to $ \sim 2\cdot10^{10}$ km$^{2}$ and a growth until a maximum is reached. The maximum phase settles at an area of $ \sim 9\cdot10^{10}$ km$^{2}$ around June 3rd 2012 and lasts about one month. In the decaying phase the area drops first sharply then moderately to $ \sim 1\cdot10^{10}$ km$^{2}$ until the CH cannot be observed anymore in October, 2012. The error bars represent the $1\sigma$ deviation from the averaging as described in Subsection~\ref{sec:methods}. The 3 phases are marked as color bar at the bottom of Figure~\ref{fig:3-mag-area-plot}(a). 

Compared to that, we plot in Figure~\ref{fig:3-mag-area-plot}(a) the evolution of the signed magnetic field strength (blue line) representing a measure of the open field within the CH, and the unsigned field strength (red line) measuring the absolute field strength. From this we see a synchronized evolutionary behavior that seems to be linked to the evolution of the CH area, nevertheless some differences are obtained. The mean field strength varies between $-1$ and $-5$~G, hence, the predominant negative polarity does not change over the CH lifetime. The profile shows one early peak ($-2.7$~G) around April 09, 2012, that does not match with the CH evolution and a main peak ($-4.4$~G) around June 03, 2012 that coincides with the peak in the area. During the maximum phase the signed mean magnetic field strength declines to $-3.8$~G before dropping significantly at the start of the decaying phase. In the decaying phase the value drops below $-1$~G. The unsigned mean magnetic field strength, shows a similar behavior to the signed field strength. It reveals an early peak ($6.7$~G) during the CH area growing phase, as well a clear maximum ($8.6$~G) and a decrease to $5.6$~G that both match the CH evolution.
 

Figure~\ref{fig:3-mag-area-plot}(b) explores the relation of the signed field strength to the area. We find a linear correlation with a Pearson correlation coefficient of $c=-0.82$ with a $95\%$ confidence interval of $[-0.36,-0.97]$. The linear regression fit for this CH can be expressed through
\begin{equation}
A  =(-1.21 \pm 0.33)+(-2.46 \pm 0.16)\cdot\bar{B}.
\end{equation}

The area, $A$ is given in $10^{10}~\mathrm{km}^{2}$ and the mean magnetic field strength $\bar{\mathrm{B}}$ is given in Gauss. A similar significant relation can be found when comparing the unsigned mean magnetic field strength to the CH area (Fig.~\ref{fig:3-mag-area-plot}(c)), where we find a linear correlation with a Pearson correlation coefficient of $c=0.83$ in a $95\%$ confidence interval of $[0.38,0.97]$. The linear regression fit can be described as
\begin{equation}
A  =(-14.22 \pm 1.25)+(2.88 \pm 0.19)\cdot\bar{B_{\mathrm{us}}}.
\end{equation}

\subsection{Magnetic Field Distribution} \label{subsec:res:distr}

\begin{figure*}
\centering \includegraphics[height=1\linewidth, angle=90]{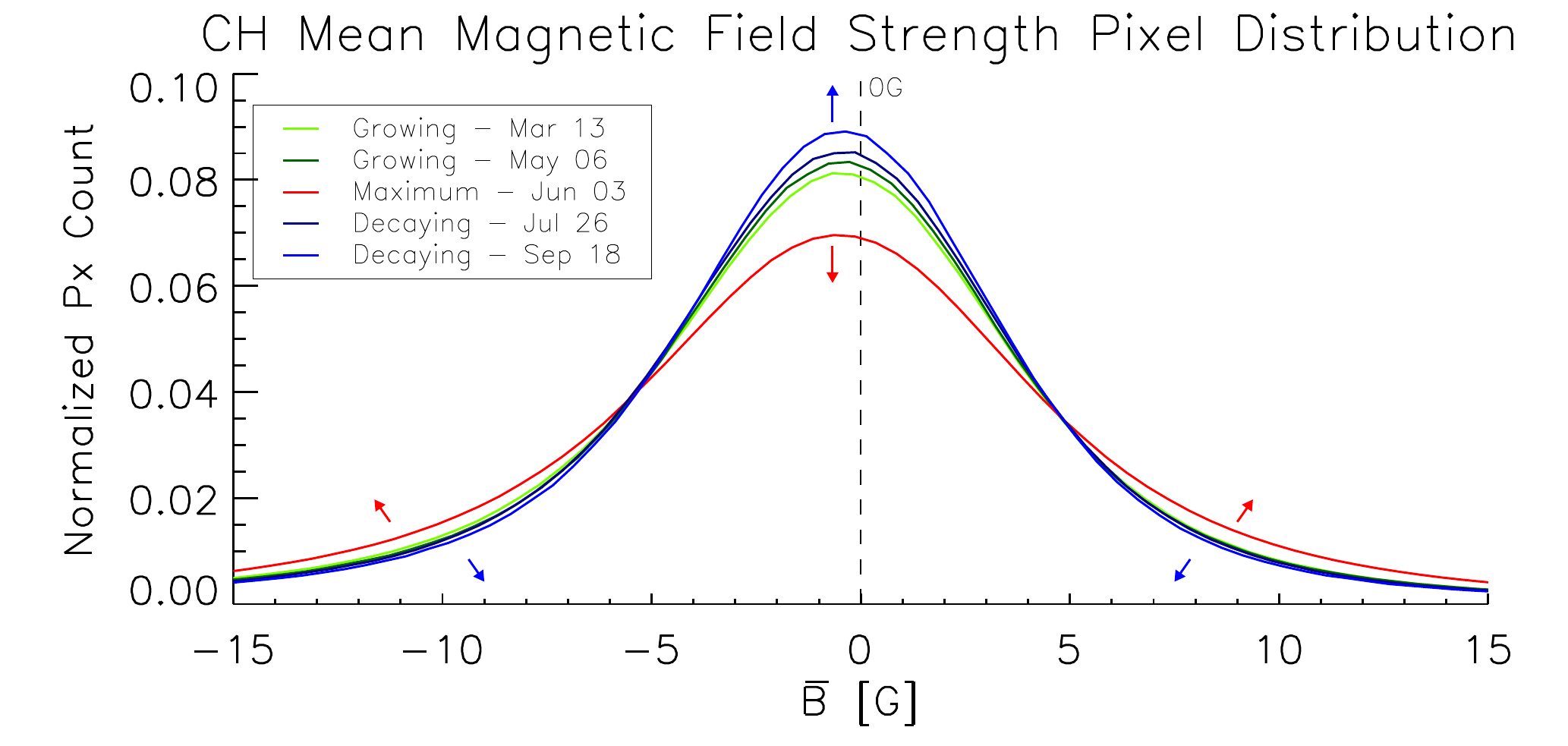}
\caption{Probability distribution of magnetic field strength of all pixels within the CH. The different lines represent different stages in the evolution of the CH. The green lines (green, dark green) represent the growing phase around March 13, and May 06, 2012. The red line is the magnetic field distribution during the maximum around June 03, 2012. The blue lines (blue, dark blue) represent the decaying phase around July 26, and  September 18, 2012.}\label{fig:mag-distr}
\end{figure*}

Figure~\ref{fig:mag-distr} shows the normalized magnetic field pixel distribution for five time steps, each corresponding to one data point in Figure~\ref{fig:3-mag-area-plot}, representing different stages in the evolution of the CH. The first two lines (green, dark green) mark the growing phase around March 13, and May 06, 2012. The red line is the magnetic field distribution during the maximum around June 03, 2012. The blue lines (blue, dark blue) represent the decaying phase around July 26, and September 18, 2012. The distribution follows a Lorentzian-like profile that is shifted to the dominant polarity of the CH (between $-0.25$~G and $-0.5$~G). By comparing the distributions of the different evolutionary stages we find a broadening from the growing to the maximum phase. We find an increase in the density of pixel with higher field strengths (flanks of the distribution) in the maximum phase compared to the growing and decaying phase. 

\begin{figure}
\centering \includegraphics[height=1\linewidth, angle=90]{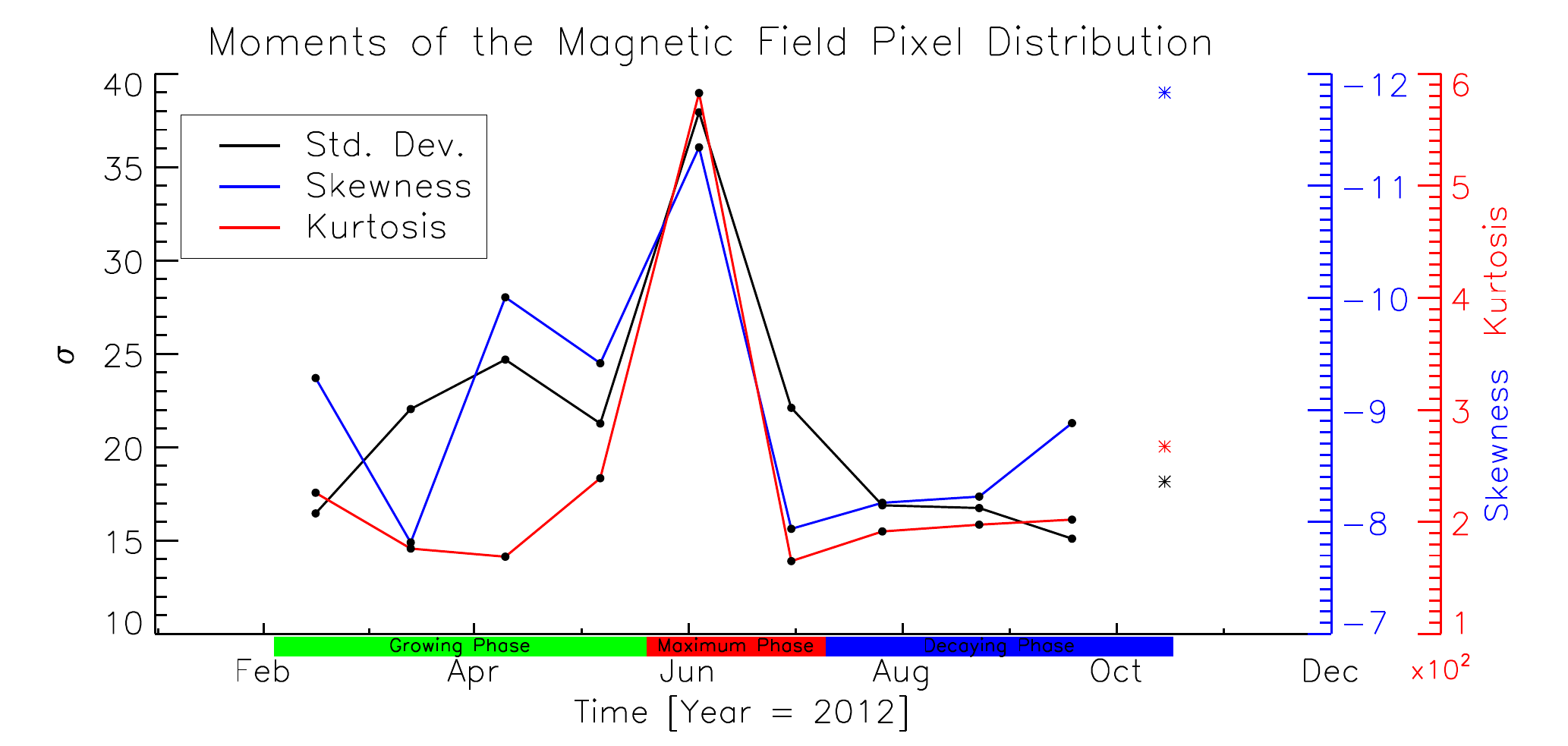}
\caption{Moments of the magnetic field distribution of the CH. The standard deviation (square root of the second moment) is represented by the black line. The blue and red lines show the skewness (third moment) and kurtosis (fourth moment) respectively. The asterisks represent the last point that was excluded due to high uncertainties (see text). }\label{fig:mom3}
\end{figure}

Figure~\ref{fig:mom3} shows the second, third and fourth moment of the magnetic field distribution, corresponding to the standard deviation (square root of the variance), the skewness (a measure of the lopsidedness of a distribution) and the kurtosis (a measure of the heaviness of the tail of the distribution). Interestingly, we find a similar behavior in all the profiles: a clear peak in the maximum phase followed by a steep drop to low values in the decaying phase. This suggests a significant change in the magnetic field distribution in the maximum phase which coincides with the maximum in the area. The asterisks represent the last datapoint (around October 14, 2012) which we excluded as an outlier in the calculated moments of the distribution, because of high uncertainties in the extraction and deprojection of the magnetic field with a small area. The outlier value arises from a few erroneously detected CH pixels, which cover strong fields. Due to the small CH area at this time, these erroneous pixels strongly alter the calculated moments of the distribution.


\subsection{Magnetic Flux} \label{subsec:res:mag-flux}
\begin{figure}
\centering \includegraphics[height=1\linewidth, angle=90]{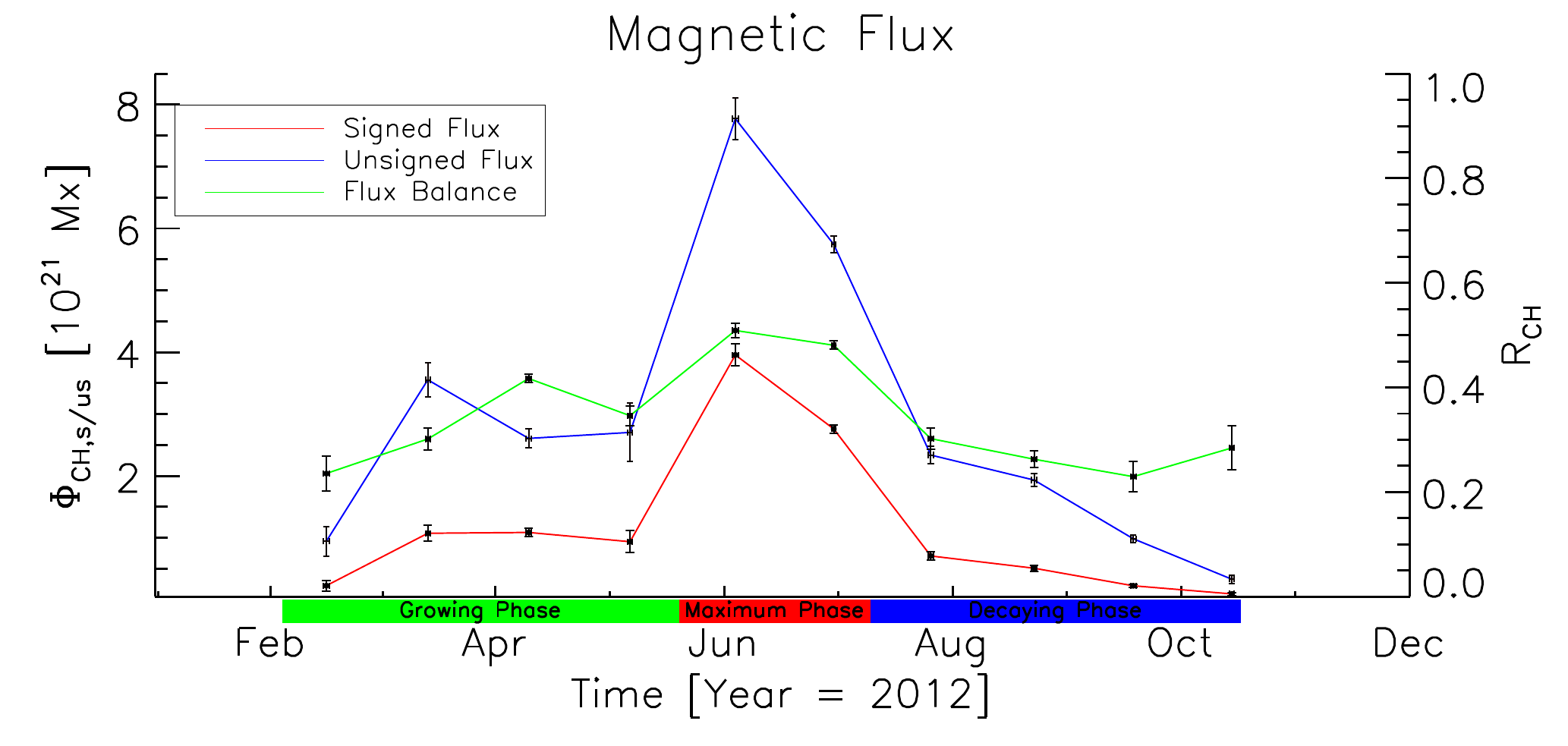}
\caption{The signed (red) and unsigned (blue) magnetic flux and their ratio, the magnetic flux balance (green). Note, the absolute value of unsigned flux is plotted.}\label{fig:flux}
\end{figure}

Figure~\ref{fig:flux} shows the CHs signed flux $\Phi_{\mathrm{s}}$, unsigned flux $\Phi_{\mathrm{us}}$ and the flux balance ($\Phi_{\mathrm{s}} / \Phi_{\mathrm{us}}$). The unsigned flux has a negative polarity but for visualization purposes in Figure~\ref{fig:flux} the absolute value is plotted. The fluxes show the same trend as the area and the mean magnetic field strength, with a growing, a maximum and a decaying phase. The signed flux (red) peaks at a maximum of $4\cdot 10^{21}$~Mx and reaches down to $1\cdot 10^{21}$~Mx during the CH "formation" and decay phase. The unsigned flux (blue) peaks at the same time as the signed flux with $8\cdot 10^{21}$~Mx and ranges down to $1.5\cdot 10^{21}$~Mx. The flux balance also peaks during the maximum phase with a flux balance of 0.5 and ranging down to 0.2 during the early and late phases. The variation within the CH area correlates with the variation of magnetic flux and clearly shows the 3-phase evolution.

\section{Flux Tube Properties} \label{sec:ft_prop}
In contrast to the global magnetic characteristics of the photospheric field covered by the CH that are presented in Section~\ref{sec:global_res}, here we analyze the magnetic fine structure within the CH, which is known to be clustered in flux tubes, magnetic small-scale structures of unipolar flux \citep{2005tu,2017hofmeister}.
 
\subsection{Flux Tube Number and Area Proportion} \label{subsec:ft_prop:num+area}
\begin{figure}
\centering \includegraphics[height=1\linewidth, angle=90]{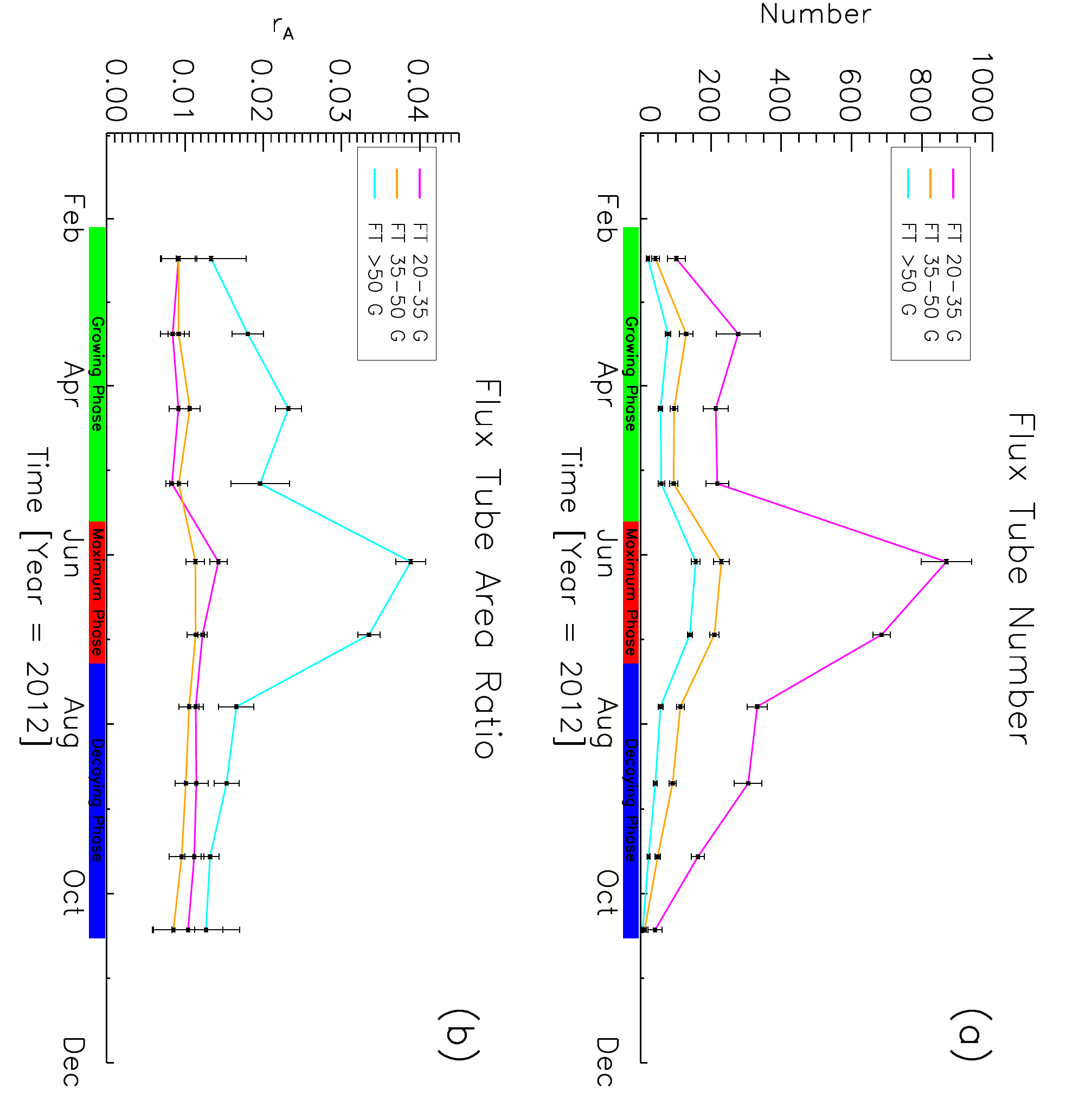}
\caption{Evolution of the number of FTs (a) and of the ratio of the summed FT area and total CH area (b) in different FT categories, represented by different colors: magenta: weak FTs; orange: medium FTs; cyan: strong FTs.}\label{fig:ft-number}
\end{figure}

\begin{figure}
\centering \includegraphics[height=1\linewidth, angle=90]{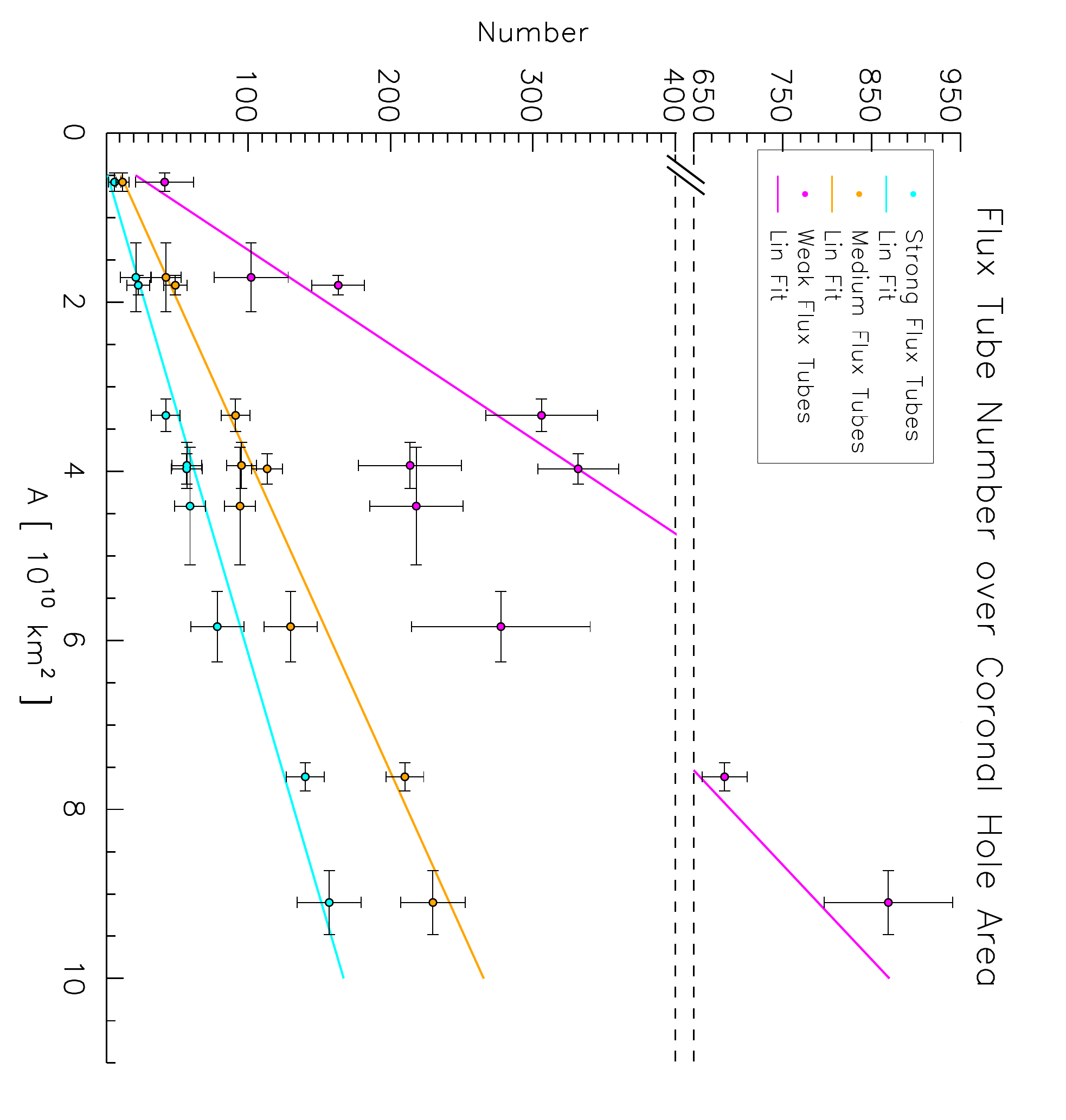}
\caption{Correlation between the number of flux tubes and the coronal hole area. The colors represent the different flux tube categories: magenta: weak FTs; orange: medium FTs; cyan: strong FTs. }\label{fig:ft-number-area}
\end{figure}

Figure~\ref{fig:ft-number}(a) shows the evolution of the average number of FTs in the CH separately for the three different FT categories, namely weak (magenta), medium (orange) and strong FTs (cyan). The number of strong FTs ranges from $7$ to over $160$, for medium FTs from $12$ to $230$ and for the weak FTs from $40$ to $870$. Although during the whole evolution, weak FTs appear most frequently, the strong FTs have the highest area proportion, i.e.\,the area of all FTs of one category in comparison to the CHs total area (Fig.~\ref{fig:ft-number}(b)). During the maximum phase, strong FTs cover up to $3.9\%$ of the total CH area. During growing and decaying phase this number is up to a factor of $3$ lower. Variations in the area proportion of the other FT categories (weak, medium) are barely given, although their number does vary slightly following the 3-phase evolution. Their area proportion averages around $(1.0 \pm 0.3)\%$. Figure~\ref{fig:FT-example} illustrates the evolution of the FT number and area proportion in 3 snapshots (one of each phase). Note, that these findings also imply that strong FTs have the highest contribution to the total CH flux, as their field strength and their area proportion is the largest. This indicates, that strong flux tubes play a major role in the evolution of a CH. 

The number of FTs and the total area they cover show the 3-phase evolution, therefore a relation would seem reasonable. Figure~\ref{fig:ft-number-area} shows the correlation between FT number and CH area using different colors for the different FT categories (weak FTs: magenta, medium FTs: orange and strong FTs: cyan). For weak FTs, we find a Pearson correlation coefficient of $c = 0.89$ with a $95\%$ confidence interval of [0.62,0.99]. Though this reveals a high correlation, however we note that the linear fit does not cover three data points (including error bars) making the correlation less significant. The medium FTs however have a more significant correlation. The Pearson correlation coefficient is $c=0.96$ with a $95\%$ confidence interval of $[0.85,0.99]$, and the linear fit can be described by
\begin{equation}
N  =(-2.5 \pm 8.1)+(26.8 \pm 1.6)\cdot A [10^{10}~ \mathrm{km}^{2}].
\end{equation}
For the strong FTs we see the strongest correlation with a Pearson correlation coefficient of $c=0.96$ with a $95\%$ confidence interval of $[0.89,0.99]$, and the linear fit can be expressed as
\begin{equation}\label{eq:FT-strong-area}
N =(-7.1 \pm 4.7)+(17.4 \pm 0.9)\cdot A [10^{10}~ \mathrm{km}^{2}].
\end{equation}

 \subsection{Flux Tube Distribution} \label{subsec:ft_prop:dist}
Figure~\ref{fig:FT-distr} shows the magnetic field distribution of flux tubes per area of different stages in the CH evolution. The color indices match Figure~\ref{fig:mag-distr}: The first two lines (green, dark green) represent the growing phase around March 13, and May 6, 2012. The red line is the magnetic field distribution during the maximum around June 3, 2012. The blue lines (blue, dark blue) represent the decaying phase around July 26, and September 18, 2012. From this we find clear differences between the distributions of the three phases. Besides from the obvious and expected asymmetry of the distributions, due to one dominant polarity, in the growing phase the peak of the distribution is slightly lower than in the following phases. Towards the maximum phase the dominant (negative)  polarity of the distribution rises, especially the flanks (i.e.\, at high field strengths), whereas the flank of the non-dominant polarity FTs declines. We find a significant rise in the FTs $< -70$~G, which shows that strong FTs, cause the main difference between the distributions. The decaying phase is marked by a decline of the dominant polarity flank and an increase of the non-dominant polarity flank. The number of weaker FTs does not change significantly (see also Fig.~\ref{fig:ft-number}).

\subsection{Flux Tube Flux Proportion} \label{subsec:ft_prop:prop}
Figure~\ref{fig:FT-flux-prop} shows, how much flux the FTs supply to the total signed flux of the CH, which may act as a measure of the \textit{open} flux of the CH. We assume that flux from FTs of opposite polarity cancels within the CH. We find that a large fraction of the signed flux of the CH comes from FTs ($70-80\%$), which occupy less than $7\%$ of the CHs area. The major part of the flux comes from strong FTs ($48-71\%$). In the evolution of FT flux proportion, the 3-phase evolution is emphasized. We see the three distinctive phases as clearly as in the area, especially in the strong FTs. For strong FTs we have a constant flux proportion of $60\%$ in the growing phase, in the maximum phase the proportion rises to $\sim 70\%$ and in the decaying phase the proportion drops down to around $54\%$. It is also interesting to note that during the maximum phase the rise in contribution of the strong FTs is accompanied by a drop in the contribution of the medium FTs. Other than that, the contribution of the medium FTs ($6-14\%$) and the weak FTs ($2.5-4.5\%$) only plays a minor role in comparison to the strong FTs.

\section{Discussion} \label{sec:dis}
Using photospheric LoS magnetograms, we have investigated the magnetic field evolution within a long-lived low-latitude CH over its entire lifetime of more than 10 solar rotations. The 3-phase evolution as found in the CH area, intensity and solar wind parameter of the associated HSS (see paper~I) can also clearly be seen from the magnetic evolution of the CH underlying photospheric field. Small structures of unipolar field (strong FTs) are derived to be the major contributer of signed flux (open magnetic field) within the CH and therefore play a key role in the magnetic evolution of the CH.


Our analysis of the CH magnetic field yields that the mean magnetic field strength of the signed and unsigned field peak at the same time as the CH area, with a maximum of $-4.4$~G and $8.6$~G respectively (Fig.~\ref{fig:3-mag-area-plot}). The evolution of the CH area and field strength are highly correlated, $c>0.80$. For the magnetic flux (Fig.~\ref{fig:flux}) we can also see the rise to a maximum of $−4 \cdot 10^{21}$~Mx (signed flux) and $8 \cdot 10^{21}$~Mx (unsigned flux). The fraction of the unbalanced (open) flux reaches a maximum of $50\%$. We find our results in good agreement with the statistical analysis of 288 low-latitude CHs by \citet{2017hofmeister}. This shows that the case study presented here is well reflected in statistical results (we note that the CH under study is a subset of the CHs used in the statistical study). 

We find an evolutionary pattern in the magnetic field distribution of the CH, with strongest changes in the flanks of the distribution. While the mean magnetic field strength, magnetic flux and area have increased values over the entire maximum phase, the distribution reveals a major change around June 3, 2012 (Fig.~\ref{fig:mag-distr},~\ref{fig:mom3}). 
We find a decrease in pixels of strong magnetic field strength ($>35$~G) in the non-dominant polarity and an increase in pixel of strong magnetic field strength of the dominant polarity. The normalized core of the magnetic field distribution however stays very constant (Lorentz-like). This time period in the evolution of the CH seems to mark the turning point in the magnetic evolution, the evolutionary peak.

\begin{figure*}
\centering \includegraphics[height=1\linewidth, angle=90]{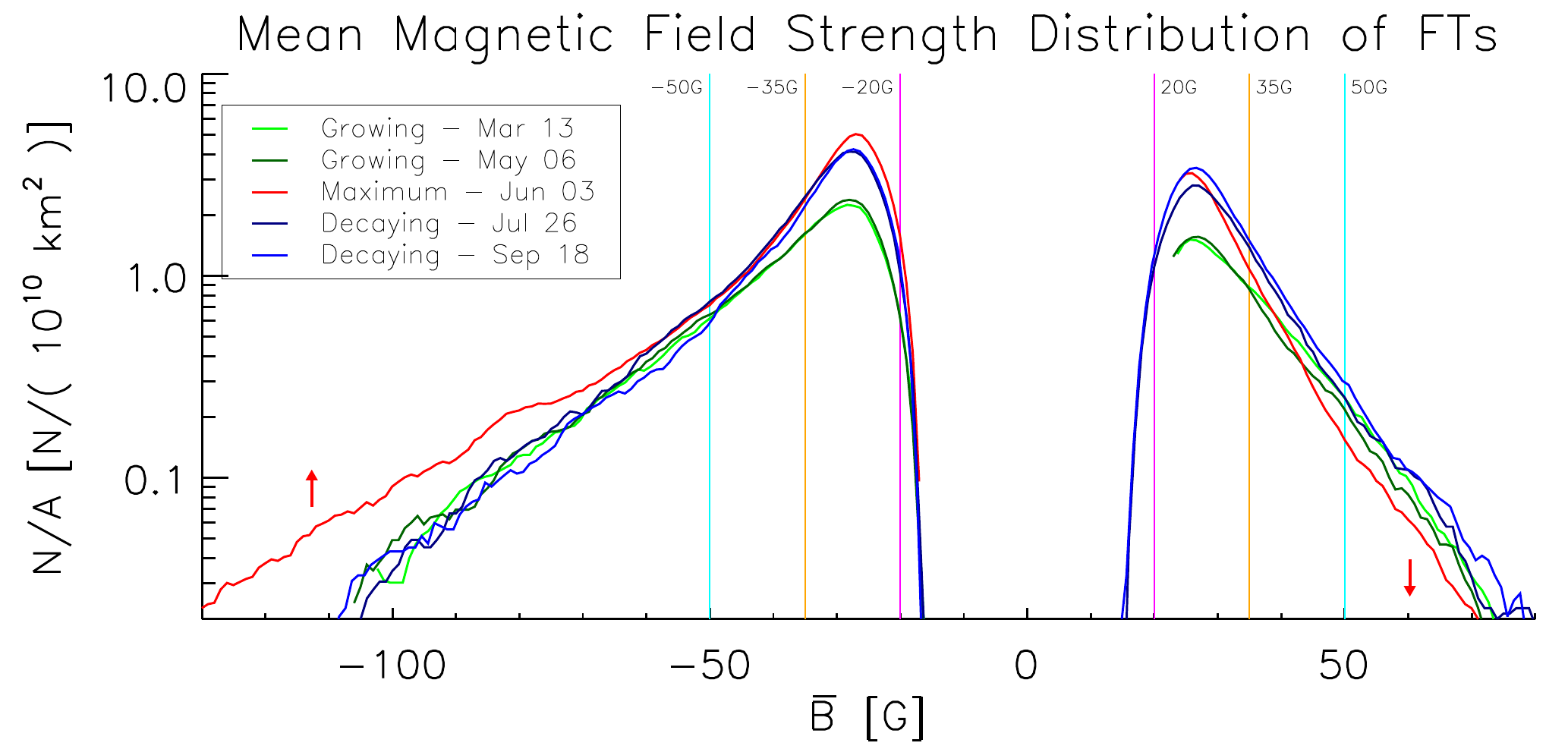}
\caption{Distribution of FTs as function of their mean magnetic field strengths during the CH evolution.}\label{fig:FT-distr}
\end{figure*}

\begin{figure}
\centering \includegraphics[height=1\linewidth, angle=90]{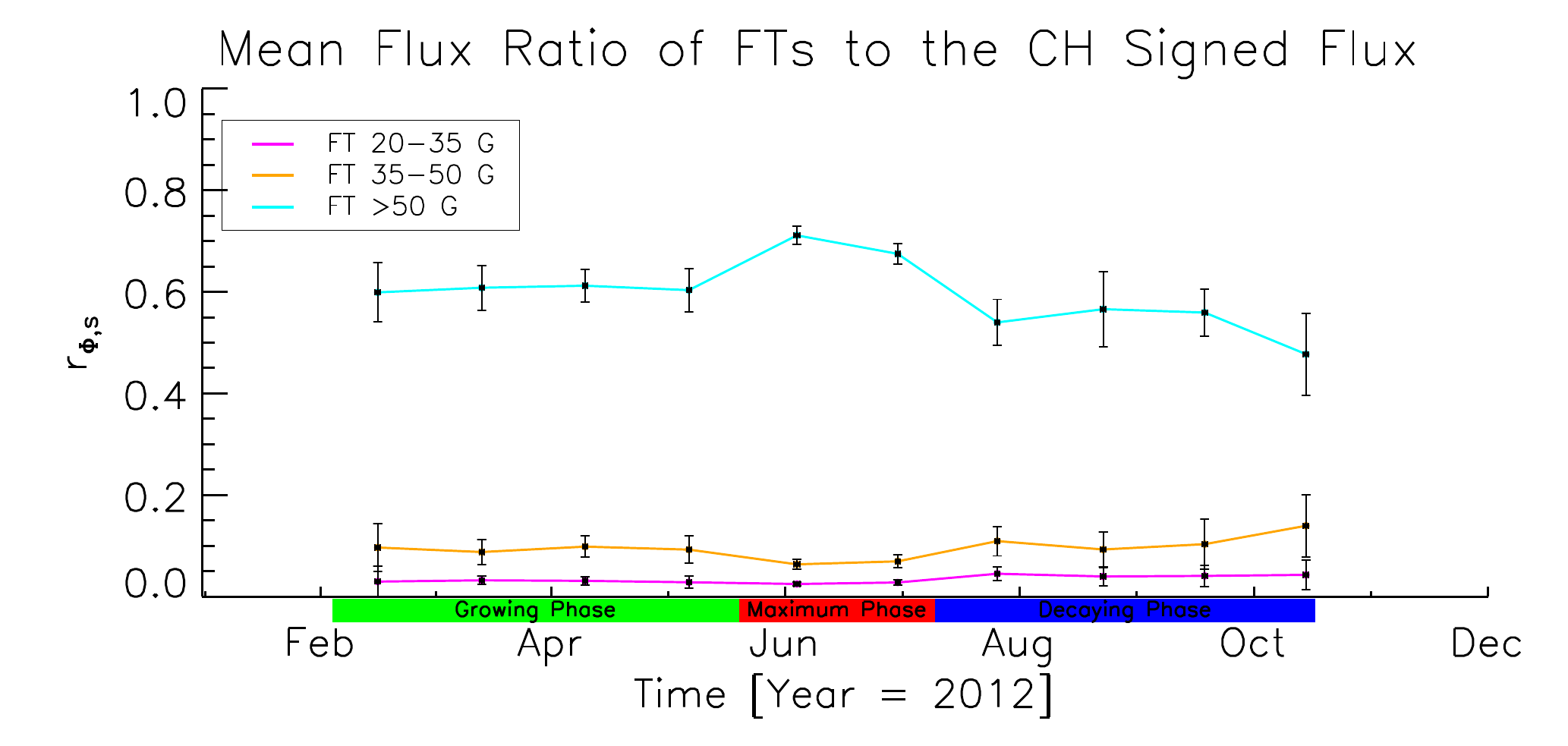}
\caption{Flux contribution of the FTs to the signed flux in the CH. The different colors represent different FT categories: magenta: weak FTs; orange: medium FTs; cyan: strong FTs. }\label{fig:FT-flux-prop}
\end{figure}

The importance of the magnetic fine structure of a CH in regards to its evolution becomes apparent when considering that FTs are the major contributor to the signed magnetic flux of the CH. Up to $80\%$ of the CHs signed flux comes from less than $7\%$ of its area. The FTs, that most likely form through flux accumulation at the edges of the magnetic network \citep{1976gabriel,1986dowdy}, expand into the corona shaping and forming the coronal structure that is observed. The FT expansion is related to the magnetic field strength \citep[e.g., ][]{wang90,2005tu,cranmer2009}, hence, changes are supposed to have a visible effect in the corona. Major changes in the distribution of the mean magnetic field strength are found for strong FTs  (cf.\,Fig.~\ref{fig:FT-distr}). Strong FTs are of highest importance as they contribute up to $70\%$ to the total signed flux of the CH, but only cover $<5\%$ of the CH area, and their evolution is closely related to the CH area evolution. A similar conclusion was drawn from a statistical study by \citet{2017hofmeister}, who found that strong FTs are the major contributor to the signed magnetic flux of a CH.

In paper I of this case study, we have shown that the CH properties are different in the growing and decaying phase of the CH. This was derived especially for the CH area and the peak bulk velocity (v$_{p}$) of the corresponding high speed stream measured in-situ at about 1 AU. In this paper we find such a behavior also for the magnetic field distribution of the CH (Fig.~\ref{fig:FT-distr}). The reason might be differences in the dynamic changes of flux emergence and solar wind plasma outflow (growing phase) as well flux cancellation (decaying phase). We suspect that the process of growing might be linked to the contribution of the internetwork to the magnetic network \citep{2014gosic}, which in CHs is usually dominated by one polarity \citep{1982levine,1996wang}. The flux is then fed into the nodes of the magnetic network, which is built up by FTs, preferably through  the merging of magnetic elements \citep{2012iida}. Also a second process, the possible induction of magnetic field caused by vortices of outflowing plasma is worth mentioning. From observations and simulations it has been shown that small scale dynamos can form and induce magnetic field (e.g. see \citealt{1993petrovay, 2008schussler, 2010graham}; reviews by \citealt{1993solanki, 2013schussler}). In both cases, an increasing field strength in the FTs results in a larger FT diameter. As FTs strive to be in a pressure equilibrium with the surrounding plasma, an increase in magnetic field strength increases the magnetic pressure which causes an expansion of the magnetic structure until the magnetic pressure and the outside gas pressure are equalized. When increasing the FT size, the catchment area for magnetic elements may also increase which would lead to a faster accumulation of flux. Consequently, an increased FT diameter may increase the possible plasma outflow. In return, this may increase the magnetic field strength. Thus, the growth of FTs (both in area and magnetic field strength) might be due to a positive feedback loop process between the magnetic field and the outflowing plasma.  

For the decaying phase, we speculate that the closing processes might either be caused by super-surface phenomena like the reconnection of open fields or/and a disruption of the positive feedback loop that grows and sustains the FTs. The efficiency of those closing processes may also be related to the configuration state of the magnetic field structure within the CH, meaning that different configurations facilitate growing and decaying processes differently. For example, if the continuous outflow in some funnels is disrupted and starts to weaken in a non-uniform or unsynchronized manner and at a higher rate than the increase in other funnels, we would expect a non-linear relation between CH area and solar wind outflow speed (which has been shown statistically, e.g. by \citealt{temmer18}).

The growth and decay of the overall CH might be related to the actual magnetic and gas pressure in the CH which is strongly determined by the properties of the FTs as compared to the ambient corona. The total pressure gradient in the CH and in particular at the boundary of the CH could lead to an effective expansion, respectively decay of the CH. The evolution of the mean magnetic field strength is shown in Figure~\ref{fig:3-mag-area-plot} and the mean intensity, which can be taken as a proxy for the gas pressure (as the intensity if determined by the density and temperature of the emitting plasma), is shown in paper I, Figure 8.

Our results not only increase our understanding of the evolution of a CH, but give valuable insights in their structure and may help to improve CH models, especially in combination with the modeling of the fast solar wind. The FTs (or their photospheric footpoints) may serve as boundary conditions to flux tube based coronal and solar wind models (e.g. see \citealt{2005tu,2014woolsey,2016pinto,2017pinto}). 


\section{Conclusions} \label{sec:con}
We have investigated the magnetic evolution of a chosen long-lived low-latitude coronal hole. The major findings can be summarized as follows:

 \begin{enumerate}


 \item We find a strong correlation between the CH area and the magnetic field strength that persists over the evolution of the CH, which shows the 3-phase evolution proposed in \citet{2018heinemann_paperI}. The signed magnetic field strength increases during the growing phase of the CH from $-2$~G up to a peak of $-4.4$~G in the maximum phase. The peak corresponds with the peak in the CH area. The decaying phase is determined by a steep drop to below $-2$~G. The unsigned mean magnetic field strength shows a nearly identical behavior with a maximum of $8.6$~G and below $6$~G in the growing and decaying phases.

\item At the maximum of the CH area as well as at the magnetic field strength and flux maximum we can derive a turning point of the evolution from the magnetic field distribution of the CH. The moments of the distributions also show a maximum there. We find an asymmetrical, Lorentzian-like profile which changes significantly in its flanks, with the core staying nearly unaffected. 

\item We find that the magnetic field of a CH is composed of FTs that are linked to the magnetic network. Within FTs, the strong ones ($>|50\mathrm{G}|$) dominate the total magnetic flux of the CH. These FTs contribute $48-71\%$ to the total signed flux of the CH, despite only covering less than $5\%$ of the CH area. The percentage contribution of signed magnetic flux ($\approx 80\%$) and area ($\approx 7\%$) from FTs to the total CH signed (open) magnetic flux and area is maximal during the maximum phase of the CH evolution.  

 \end{enumerate}


Our case study of a long-lived CH clearly shows that magnetic FTs are the elementary building blocks governing the CH evolution. Notably, during its maximum phase, the distribution of strong FTs of the dominant polarity reveal a strong intensification and govern the overall magnetic CH characteristics. Comparing these findings to the results in paper I, we also note that strong changes in the magnetic fine structure of the CH are reflected in a change of the characteristics of the associated high speed solar wind streams.

\section*{Acknowledgements}
The SDO/AIA and SDO/HMI image data is available by courtesy of NASA/SDO and the respective science teams. We acknowledge the support by the FFG/ASAP Program under grant no. 859729 (SWAMI). A.M.V.\,and M.T.\,acknowledge the Fonds zur F\"orderung wissenschaftlicher Forschung (FWF): P24092-N16 and V195-N16. S.J.H.\,acknowledges support from the JungforscherInnenfonds der Steiermärkischen Sparkassen.

\appendix

\startlongtable
\begin{deluxetable}{l c |c c c c c}
\tablecaption{Correlation Coefficients Overview\label{tab:corr}}
\tablehead{
\colhead{} & \colhead{}& \multicolumn{5}{c}{Pearson Correlation Coefficient}  \\
\colhead{Relation} &\colhead{Figure Nr.}& \colhead{$\mu_{\mathrm{P}}$} & \colhead{$\sigma_{\mathrm{P}}$} & \colhead{CI $90\%$} &\colhead{CI $95\%$} &\colhead{CI $99\%$}
} 
\startdata
$\bar{B}_{\mathrm{s}}$ vs. $A_{\mathrm{CH}}$ 	& \ref{fig:3-mag-area-plot}(b) & $-0.82$ & $0.15$ & $[-0.48,-0.96]$ & $[-0.37,-0.97]$ & $[-0.15,-0.99]$   \\
$\bar{B}_{\mathrm{us}}$ vs. $A_{\mathrm{CH}}$ &\ref{fig:3-mag-area-plot}(c)  &  $0.84$ & $0.14$ & $[0.54,0.97]$ & $[0.42,0.97]$ & $[0.20,0.99]$   \\
$A_{\mathrm{CH}}$ vs. $N_{\mathrm{FT,strong}}$ & \ref{fig:ft-number-area} &  $0.96$ & $0.03$ & $[0.91,0.99]$ & $[0.89,0.99]$ & $[0.83,1.00]$   \\ 
$A_{\mathrm{CH}}$ vs. $N_{\mathrm{FT,medium}}$ & \ref{fig:ft-number-area} &  $0.96$ & $0.04$ & $[0.88,0.99]$ & $[0.85,0.99]$ & $[0.75,1.00]$   \\ 
$A_{\mathrm{CH}}$ vs. $N_{\mathrm{FT,weak}}$ & \ref{fig:ft-number-area} &  $0.89$ & $0.10$ & $[0.71,0.98]$ & $[0.62,0.99]$ & $[0.35,1.00]$    \\ \hline \hline
\multicolumn{2}{ c }{ } & \multicolumn{5}{ c }{Spearman Correlation Coefficient}\\
\multicolumn{1}{ c }{Relation}&\multicolumn{1}{ c }{Figure Nr. } & $\mu_{\mathrm{S}}$ &$\sigma_{\mathrm{S}}$ & CI $90\%$ &CI $95\%$ &CI $99\%$ \\ \hline
$\bar{B}_{\mathrm{s}}$ vs. $A_{\mathrm{CH}}$ &\ref{fig:3-mag-area-plot}(b)  &$-0.73$ & $0.18$ & $[-0.38,-0.94]$ & $[-0.27,-0.95]$ & $[-0.06,-0.98]$\\
$\bar{B}_{\mathrm{us}}$ vs. $A_{\mathrm{CH}}$&\ref{fig:3-mag-area-plot}(c) &$0.74$ & $0.17$ & $[0.41,0.95]$ & $[0.31,0.96]$ & $[0.09,0.99]$\\
$A_{\mathrm{CH}}$ vs. $N_{\mathrm{FT,strong}}$ & \ref{fig:ft-number-area} &$0.91$ & $0.08$ & $[0.76,0.99]$ & $[0.70,0.99]$ & $[0.54,1.00]$\\ 
$A_{\mathrm{CH}}$ vs. $N_{\mathrm{FT,medium}}$ & \ref{fig:ft-number-area} &  $0.88$ & $0.11$ & $[0.67,0.98]$ & $[0.60,0.99]$ & $[0.39,1.00]$   \\ $A_{\mathrm{CH}}$ vs. $N_{\mathrm{FT,weak}}$ & \ref{fig:ft-number-area} &  $0.78$ & $0.18$ & $[0.43,0.96]$ & $[0.32,0.98]$ & $[0.06,0.99]$ \\  \hline 
\enddata
\end{deluxetable}


\bibliographystyle{aasjournal}

\end{document}